\documentclass[12pt,prd,tightenlines,nofootinbib,showpacs,showkeys]{revtex4}
\newcommand{\be}{\begin{equation}}
\newcommand{\ee}{\end{equation}}
\newcommand{\bdis}{\begin{displaymath}}
\newcommand{\edis}{\end{displaymath}}
\newcommand{\bga}{\begin{equation}\begin{gathered}}
\newcommand{\ega}{\end{gathered}\end{equation}}
\usepackage{bm}
\usepackage{graphics}
\usepackage{rotating}
\usepackage{epsfig}
\usepackage{amssymb}
\usepackage{amsmath}
\usepackage{amsthm}
\usepackage{amsfonts}
\usepackage{amssymb}
\usepackage{braket}
\usepackage{graphics}
\usepackage{rotating}
\usepackage{epsfig}
\usepackage{indentfirst}
\usepackage{epstopdf}

\begin{document}
\title{\begin{flushright}{\rm\normalsize SSU-HEP-15/11\\[5mm]}\end{flushright}
Hadronic deuteron polarizability contribution\\
to the Lamb shift in muonic deuterium}
\author{\firstname{A.~V.} \surname{Eskin}}
\affiliation{Samara State Aerospace University named after S.P. Korolyov, Moskovskoye Shosse 34, 443086,
Samara, Russia}
\author{\firstname{R.~N.} \surname{Faustov}}
\affiliation{Dorodnicyn Computing Centre, Russian Academy of Science, Vavilov Str. 40, 119991, Moscow, Russia}
\author{\firstname{A.~P.} \surname{Martynenko}}
\affiliation{Samara State University, Pavlov Street 1, 443011, Samara, Russia}
\affiliation{Samara State Aerospace University named after S.P. Korolyov, Moskovskoye Shosse 34, 443086,
Samara, Russia}
\author{\firstname{F.~A.} \surname{Martynenko}}
\affiliation{Samara State University, Pavlov Street 1, 443011, Samara, Russia}

\begin{abstract}
Hadronic deuteron polarizability correction to the Lamb shift of muonic deuterium is calculated
on the basis of unitary isobar model and modern experimental data on the structure functions of deep
inelastic lepton-deuteron scattering and their parameterizations in the resonance and nonresonance regions.
\end{abstract}



\maketitle

Finding more accurate values of the charge radii of the proton and deuteron in experiments of the CREMA collaboration
\cite{CREMA,CREMA2015,CREMA1} requires a comparison of experimental data and theoretical calculations of the Lamb shift.
The deuteron structure and polarizability contributions to the Lamb shift $(2P-2S)$ and hyperfine structure in
muonic deuterium are among that corrections which are not known at present with
the accuracy comparable with pure electromagnetic corrections \cite{egs,p1,p2,p3}. The deuteron polarizability effect
is related with the excitation of particles which enter in its composition. Hadronic deuteron
polarizability is determined by numerous nuclear reactions of $\pi$-, $\eta$-meson production on proton or neutron
and the excitation of nucleon resonances.
Earlier studies on the contribution of the nucleus polarizability in the hyperfine structure and the Lamb shift
showed that this contribution has an important role to achieve high accuracy of calculation \cite{pol1,pol2,pol4,pol5,pol6,pol7}.
(see other numerous references in \cite{egs}).
During last ten years experimental study of electromagnetic excitations of baryonic resonances was carried out
at CLAS detector (CEBAF Large Acceptance Spectrometer) \cite{mokeev}. It allowed to improve values of phenomenological
parameters describing low-energy photon nucleon interaction, helicity amplitudes $A_{1/2}$, $A_{3/2}$,
$S_{1/2}$ for different states $N^\ast$, the cross sections, angular distributions
for final states of particles. Nuclear structure corrections including effect of the deuteron polarizability were
investigated in \cite{carlson,kp2015} by means of modern parameterizations of the deuteron virtual
photoabsorption data. The aim of this work is to present another calculation of the deuteron
polarizability contribution to the Lamb shift on the basis of model MAID \cite{MAID} and modern experimental data
on the deuteron structure functions $F_2^d$ and $R^d$.

High energy charged lepton scattering is a well-established tool to investigate the structure of the deuteron.
The contribution of hadronic deuteron polarizability of order $(Z\alpha)^5$ to the Lamb shift is determined
by the amplitude of virtual forward Compton scattering $\gamma^\ast+d\to\gamma^\ast+d$ which contains the deuteron tensor \cite{RF,JB}:
\begin{equation}
\label{eq:1}
M_{\mu\nu}^{(d)}=\left(-g_{\mu\nu}+\frac{k_\mu k_\nu}
{k^2}\right)C_1(\nu,k^2)+\frac{1}{m^2_d}\left(p_{2 \mu}-\frac{m_d\nu}{k^2}k_\mu\right)
\left(p_{2 \nu}-\frac{m_d\nu}{k^2}k_\nu\right)C_2(\nu,k^2)+
\end{equation}
\begin{displaymath}
+i\varepsilon_{\mu\nu\alpha\beta}k^\alpha\Biggl\{m_dS^\beta H_1(\nu,k^2)+[(P\cdot k)S^\beta-(S\cdot k)P^\beta]
\frac{H_2(\nu,k^2)}{m_d}\Biggr\},
\end{displaymath}
where $k$ is the four-momentum of the virtual photon, $\nu=k_0$ is the virtual
photon energy, $m_d$ is the deuteron mass, $S$, $P$ are the deuteron spin and momentum four-vectors.
The deuteron tensor amplitude $M_{\mu\nu}$ consists of a spin independent symmetric part
which is defined by the $C_1(\nu,k^2)$ and $C_2(\nu,k^2)$ unpolarized structure functions, and a spin-dependent
part which is defined by the $H_1(\nu,k^2)$ and $H_2(\nu,k^2)$ spin structure functions.
Symmetrical part of the tensor \eqref{eq:1} gives the contribution to the Lamb shift (structure functions
$C_{1,2}(\nu,k^2)$) and antisymmetric part contributes to the hyperfine
structure (structure functions $H_{1,2}(\nu,k^2)$).
The structure functions $C_i(k_0, k^2)$ obey the following dispersion relations \cite{drechsel2003}:
\begin{equation}
\label{eq:2}
C_1(k_0,k^2)=C_1(0,k^2)+\frac{1}{\pi}k_0^2\int_{\nu_0}^\infty\frac{d\nu^2}
{\nu^2(\nu^2-k_0^2)}Im C_1(\nu, k^2),
\end{equation}
\begin{equation}
\label{eq:3}
C_2(k_0,k^2)=\frac{1}{\pi}\int_{\nu_0}^\infty\frac{d\nu^2}
{(\nu^2-k_0^2)}Im C_2(\nu, k^2),
\end{equation}
\begin{displaymath}
\nu_0=m_\pi+\frac{1}{2m_d}(Q^2+m_\pi^2),~~Q^2=-k^2.
\end{displaymath}
The exchanged virtual photon transfers four-momentum $k$ with the virtuality $Q^2=-k^2$. At $Q^2\gg 1~GeV^2$
deep inelastic scattering resolves the partonic constituents (quarks and gluons) of the nucleon. At $Q^2\le 1~GeV^2$
the excitation of nucleon resonances and multi-pion continuum states is important.
The threshold value of the photon energy $\nu_0$ represents the minimal
energy needed for the production of the $\pi$-meson in the reaction
$\gamma^\ast+d\to\pi^0+d$. When the energy transfer in the scattering process increases beyond the point
corresponding to the pion production threshold (i.e. when the combined invariant mass of the exchanged
virtual photon and the target exceeds the value $W_\pi = M_N +m_\pi = 1.072$ GeV),
we leave the region of elastic scattering and enter the region of inelastic scattering.
Let us note that reliable data on the
subtraction term in the first dispersion integral \eqref{eq:2} are absent. But in
the limit of small values of $k^2$ this term is related with the proton
magnetic polarizability:
\begin{equation}
\label{eq:4}
\lim_{k^2\rightarrow 0}\frac{C_1(0,k^2)}{k^2}=\frac{m_d}{\alpha}\beta_M^d.
\end{equation}
For the case of hadronic deuteron polarizability we take $\beta_M^d=\beta_M^p+\beta_M^n=6.20(2.04)\times 10^{-4}~fm^3$ \cite{PDG},
that is as the sum of the magnetic polarizabilities of the proton and neutron.
In the approximation, which is then used for the calculation the deuteron appears as a loosely coupled system
of the proton and neutron.
The dipole parameterization for the function $\beta_M(k^2)$ was suggested in \cite{kp1999}:
\begin{equation}
\label{eq:5}
\beta_M(k^2)=\beta_M^d\frac{\Lambda^8}{(\Lambda^2+k^2)^4},
\end{equation}
where $\Lambda^2=0.71~GeV^2$ is taken as for elastic nucleon form
factor. Imaginary parts of the amplitudes $C_i(k_0,k^2)$ are
expressed in terms of the structure functions $F^d_i(x,Q^2)$ of deep inelastic scattering as follows:
\begin{equation}
\label{eq:6}
\frac{1}{\pi}Im C_1(\nu,Q^2)=\frac{F^d_1(\nu,Q^2)}{m_d},~~~\frac{1}
{\pi}Im C_2(\nu,Q^2)=\frac{F^d_2(\nu,Q^2)}{\nu}.
\end{equation}
Using relations \eqref{eq:3}-\eqref{eq:6} and transforming the integration in the loop
amplitudes to four-dimensional Euclidean space
we can perform the integration over the angle variables and represent
the deuteron polarizability contribution to the Lamb shift of muonic deuterium atom
in the form \cite{fm,m1,fm1}:
\begin{equation}
\label{eq:7}
\Delta E^{LS}_{pol}=-\frac{2\mu^3(Z\alpha)^5}{\pi n^3m_1^4}\int_0^\infty
d k\int_{\nu_0}^\infty dy ~{\cal F}(y,k)
+\frac{2\mu^3(Z\alpha)^4}{\pi n^3m_1}\int_0^\infty h(k^2)\beta_M(k^2)kdk,
\end{equation}
\begin{equation}
\label{eq:8}
{\cal F}(y,k)=\frac{1}{(R^d+1)st^2(4s-t)}\Biggl\{-8\sqrt{s}(1+s)^{3/2}
\sqrt{t}(2s+R^d)-
\end{equation}
\begin{displaymath}
-\sqrt{t}(t-4s)[t+s(6+2R^d+4s+t)]+\sqrt{4+t}[(t-2)t+
s\left(8+t^2+2R^d(t+4)\right)]\Biggr\} F^d_2(y,k^2),
\end{displaymath}
\begin{equation}
\label{eq:9}
h(k^2)=1+\left(1-\frac{t}{2}\right)\left(\sqrt{\frac{4}{t}+1}-1\right),~~t=
\frac{k^2}{m_1^2},~~s=\frac{y^2}{k^2},
\end{equation}
where $R^d(y,k^2)=\sigma^d_L/\sigma^d_T$ is the ratio of the absorption cross sections
longitudinally and transversely polarized photons by hadrons.
In this way, the correction for the polarizability $\Delta E^{LS}_{pol}$ can be expressed
in terms of two structure functions $F^d_2(\nu,k^2)$ and $R^d(\nu,k^2)$, describing unpolarized
lepton-deuteron scattering. General expression \eqref{eq:7} for the polarizability correction
has the same form for muonic hydrogen and muonic deuterium. The difference is related with
different structure functions $F_2^p$ and $F_2^d$ which enter in \eqref{eq:7} and the reduced mass
dependence. Thus, to obtain the numerical value of the correction on the deuteron
polarizability, we can use the experimental data on deuteron structure functions
$F^d_{1,2}(\nu,k^2)$ and their different parameterizations \cite{deut1,deut2,deut3}.

The largest contribution to \eqref{eq:7} is determined by the region of the variable $k^2$: $0 \div  1~GeV^2$
and near-threshold values of the photon energy $\nu$. So, the exact construction of structure
functions $F^d_2$, $R^d$ in this region is extremely important to get a reliable estimate of the effect of the deuteron
polarizability. The deuteron is weakly bound system, therefore, considering the amplitude of photon-deuteron interaction,
we can assume that a photon interacts with free proton or neutron. The second nucleon plays in this case the role of the spectator.
For the deuteron no essential medium modifications are expected because of the weak binding of the nucleons.
It is assumed also that unpolarized inclusive charged-lepton-nucleon scattering $l+N\to l'+X $
(where $X$ denotes the undetected final state) occurs due to single-photon interaction.
An important kinematical variable of this process is the mass of the undetected hadronic system $W$:
\begin{equation}
\label{eq:10}
W^2=m_N^2-Q^2+2m_N\nu,~~k^2=-Q^2,
\end{equation}
where $m_N$ is the nucleon mass (proton or neutron). Using the variable \eqref{eq:10} we can divide total
integration region in \eqref{eq:7} on the resonance region $W\leq 2$ GeV where the production of low-lying nucleon
resonances occurs and deep inelastic region when $W > 2$ GeV.

There exists several possibilities to estimate the polarizability contribution \eqref{eq:7} in the resonance region.
Let us briefly characterize these approaches. First approach
is related with the use of isobar model describing photo- and electro-production of $\pi$-, $\eta$-mesons and nucleon
resonances in the $\gamma^\ast d$ interaction.
In the considered region of the variables $k^2$, $W$ the most
significant contribution
is given by five resonances: $P_{33}(1232)$, $S_{11}(1535)$, $D_{13}(1520)$,
$P_{11}(1440)$, $F_{15}(1680)$.
A coherent decomposition of the one $\pi$-meson and one $\eta$-meson production cross sections on the nucleon into resonance
and background contributions is necessary because interference terms are quite important.
In the isobar model the resonance amplitudes are parameterized in the Breit-Wigner form.
Then accounting the resonance decays to the $N\pi$- and $N\eta$-states the resonance cross
sections for the absorption of transversely polarized photons have the form \cite{Walker, Arndt,T1,T2,KAA,Bianchi,UIM1,Dong1,T3}:
\begin{equation}
\label{eq:11}
\sigma_{1/2,3/2}^T=\left(\frac{k_R}{k}\right)^2\frac{W^2\Gamma_\gamma\Gamma_{R
\rightarrow N\pi}}{(W^2-M_R^2)^2+W^2\Gamma_{tot}^2}\frac{4m_N}{M_R\Gamma_R}
|A_{1/2,3/2}|^2,
\end{equation}
where ${A_{1/2,3/2}}$ are the transverse helicity amplitudes, and
\begin{equation}
\label{eq:12}
\Gamma_\gamma=\Gamma_R\left(\frac{k}{k_R}\right)^{j_1}\left(\frac{k_R^2+X^2}
{k^2+X^2}\right)^{j_2}.
\end{equation}
The width of one-pion decay of the resonance is parameterized as follows:
\begin{equation}
\label{eq:13}
\Gamma_{R\rightarrow N\pi}(q)=\Gamma_R\frac{M_R}{M}\left(\frac{q}{q_R}\right)^3
\left(\frac{q_R^2+C^2}{q^2+C^2}\right)^2,
\end{equation}
for the resonance $P_{33}(1232)$ and
\begin{equation}
\label{eq:14}
\Gamma_{R\rightarrow N\pi}(q)=\Gamma_R\left(\frac{q}{q_R}\right)^{2l+1}
\left(\frac{q_R^2+\delta^2}{q^2+\delta^2}\right)^{l+1},
\end{equation}
for resonances $D_{13}(1520)$, $P_{11}(1440)$, $F_{15}(1680)$. $l$ is the pion angular
momentum, $\delta^2=(M_R-m_N-m_\pi)^2+\Gamma_R^2/4$. Here $q$ $(k)$ and $q_R$ $(k_R)$
denote the c.m.s. pion (photon) momenta of resonances with the mass
$M$ and $M_R$ respectively. In the case of $S_{11}(1535)$
the $\pi N$ and $\eta N$ decay modes are taken into account:
\begin{equation}
\label{eq:15}
\Gamma_{R\rightarrow N\pi,N\eta}=\frac{q_{\pi,\eta}}{q}b_{\pi,\eta}\Gamma_R
\frac{q_{\pi,\eta}^2+C_{\pi,\eta}^2}{q^2+C_{\pi,\eta}^2},
\end{equation}
where $b_{\pi,\eta}$ is the $\pi$, $\eta$ branching ratio. Numerous parameters entering
in \eqref{eq:11}- \eqref{eq:15} were fitted to experimental data and improved after new
experiments at CLAS. The cross section
$\sigma_L$ is determined by an expression similar to \eqref{eq:11} where we should
change $A_{1/2,3/2}$ on the longitudinal amplitude $S_{1/2}$. The calculation
of helicity amplitudes $A_{1/2}$, $A_{3/2}$, $S_{1/2}$ as functions of
$Q^2$ was done on the basis of the oscillator quark model in \cite{Dong2,Isgur,CL,Capstick,LBL,Warns}.

Double pion production in the reaction $\gamma^\ast+d\to \pi+\pi+d$ is another important
process for the calculation of hadronic polarizability correction.
The two-pion decay modes of the higher nucleon resonances $S_{11}(1535)$,
$D_{13}(1520)$, $P_{11}(1440)$, $F_{15}(1680)$ are described phenomenologically
using the two-step process as in \cite{T3}. The high-lying nucleon resonance
$R$ can decay first into $N^\ast$ ($P_{33}$ or $P_{11}$) and a pion or
into a nucleon and $\rho$, $\sigma$ meson. Then new resonances decay
into a nucleon and a pion or two pions:
\begin{equation}
R\rightarrow r+a=\Biggl\{{N^\ast+\pi\rightarrow
N+\pi+\pi,\atop \rho(\sigma)+N\rightarrow N+\pi+\pi.}
\end{equation}
The two-pion decay width is then given by a phase-space weighted integral over the mass distribution of
the intermediate resonance $r$ = $N^\ast, \rho, \sigma$ ($ a=\pi, N$):
\begin{equation}
\label{eq:2pi}
\Gamma_{R\rightarrow r+a}(W)=\frac{P_{2\pi}}{W}\int_0^{W-m_a}d\mu \cdot p_f\frac{2}{\pi}
\frac{\mu^2\Gamma_{r,tot}(\mu)}{(\mu^2-m_r^2)^2+\mu^2\Gamma_{r,tot}^2(\mu)}
\frac{(M_R-m_N-2m_\pi)^2+C^2}{(W-m_N-2m_\pi)^2+C^2}.
\end{equation}
The factor $P_{2\pi}$ must be taken from the constraint condition: $\Gamma_{R\to r+a}(W_R)$
coincides with the experimental data in the resonance point. $p_f$  is the three-momentum of the resonance $r$
in the rest frame of $R$. $\Gamma_{r,tot}$ is the total width of the resonance $r$. The decay width of the
meson resonance in \eqref{eq:2pi}) is parameterized similarly to that of the $P_{33}(1232)$:
\begin{equation}
\Gamma(\mu)=\Gamma_r\frac{m_r}{\mu}\left(\frac{q}{q_r}\right)^{2J_r+1}
\frac{q_r^2+\delta^2}{q^2+\delta^2},
\end{equation}
where $m_r$ and $\mu$ are the mean mass and the actual mass of the meson resonance, $q$ and $q_r$
are the pion three momenta in the rest frame of the resonance with masses $\mu$ and $m_r$. $J_r$ and $\Gamma_r$
are the spin and decay width of the resonance with the mass $m_r$.

\begin{figure}[t!]
\centering
\includegraphics[scale=0.6]{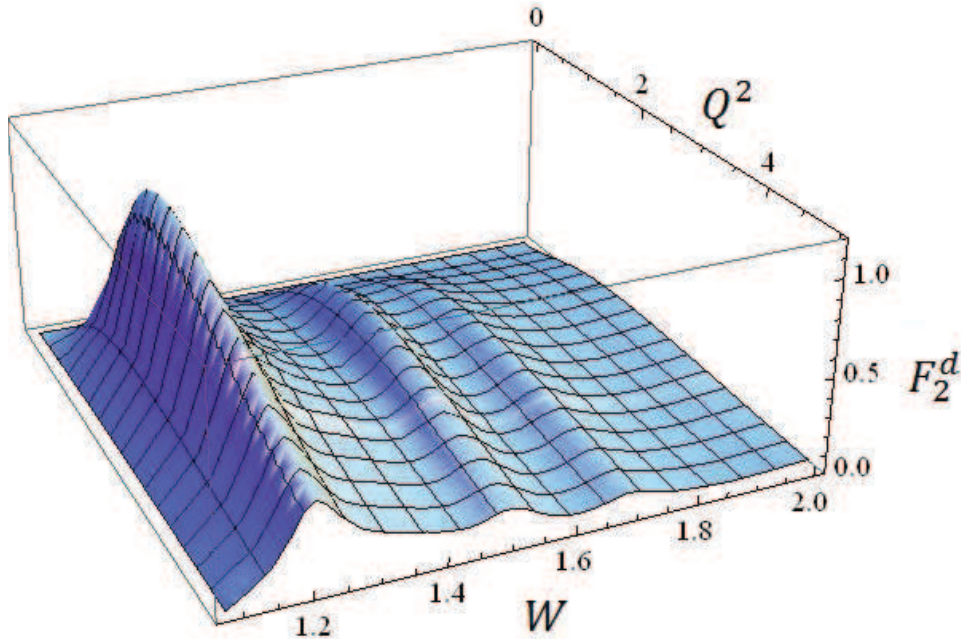}
\caption{The deuteron structure function $F_2^d$ per nucleon in $\mu b$ as a function of
$Q^2$ $(0\div 5)~GeV^2$ and $W$ $(1.1\div 2.0)~GeV$.}
\label{fig:f2d}
\end{figure}

Nonresonance contributions to the cross sections $\sigma_{T,L}$ in the resonance region are determined by the
Born terms constructed on the basis of Lagrangians of $\gamma NN$, $\gamma\pi\pi$, $\pi NN$
interactions. Another part of nonresonance background contains the $t$- channel contributions of $\rho$,
$\omega$ mesons obtained by means of effective Lagrangians $\gamma \pi V$, $VNN$ interactions ($V=\rho,\omega$)
\cite{MAID,UIM1}. In the unitary isobar model accounting the Born terms,
the vector meson, nucleon resonance contributions and the interference terms we calculated the cross sections
$\sigma^{p,n}_{T,L}$ by means of numerical program MAID (http://www.kph-uni-mainz.de/MAID) in the resonance region
as the functions of two variables $W$ and $Q^2$. After that the structure function $F^{p,n}_2(W,Q^2)$ is constructed as follows:
\begin{equation}
\label{eq:19}
F^{p,n}_2(W,Q^2)=\frac{Q^2}{4\pi^2\alpha}\left(\sigma_T+\sigma_L\right)\frac{K\nu}
{(Q^2+\nu^2)},
\end{equation}
where $K$ is the flux factor of virtual photons for which we use the Gilman definition $K_G=\sqrt{Q^2+\nu^2}$
as in \cite{fm,m1}. We obtain total cross sections for the proton and neutron
$\sigma^{p,n}_{tot}(W,Q^2)=(\sigma_T^{p,n}+\sigma_L^{p,n})$. In Fig.~\ref{fig:f2d} we present a plot of deuteron
structure function $F_2^d$ obtained by means of structure functions $F_2^p$ and $F_2^n$ in accordance with the formula
\begin{equation}
F_2^d(W,Q^2)=F_2^p(W,Q^2)+F_2^n(W,Q^2).
\label{eq:f2pf2n}
\end{equation}
It means that the photon interacts independently with a proton or a neutron. Let us note here that the authors in many papers
adopt the convention that $F_2^d$ refers to the average structure function of the nucleon in the deuteron:
$F_2^d=(F_2^p+F_2^n)/2$.
The real deuteron structure function can be represented in terms of the sum of the structure functions
of the proton and the neutron as in \eqref{eq:f2pf2n}, considering the square modulus of the full virtual Compton scattering amplitude by a deuteron.
The plot in Fig.~\ref{fig:f2d} contains three clear peaks corresponding to resonances $P_{33}(1232)$, $D_{13}(1520)$, $F_{15}(1680)$.
The theoretical construction of $F_2^d$ could be improved by the account of two-pion decays of resonances as described in \eqref{eq:2pi}.

The total value of deuteron polarizability correction is represented in Table~\ref{tb1} as the sum of the individual,
the most important contributions. We investigate contributions to the correction \eqref{eq:7} which have numerical value of order
1 $\mu eV$ for the $1S$-state in muonic deuterium. In the phenomenological model MAID we keep processes of the
$\eta$-meson production on deuterons, the production processes of the $K$-mesons, the contribution of two-pion intermediate states
in the reaction $\gamma^\ast+ d\to d+\pi+\pi$.
The basic contribution to the polarizability effect is given
by processes of the $\pi$-meson production on the deuteron in the reactions $\gamma^\ast+d\to\pi^0(\pi^+)+p+n(n+n)$,
$\gamma^\ast+d\to\pi^0(\pi^-)+p+n(p+p)$ including the resonance reactions.
The optical theorem allows us to express the imaginary part of the forward Compton scattering amplitude
through the production cross section of particles in intermediate state (see Eq.(6)).
Note, that the MAID program allows to calculate the proton and neutron cross sections $\sigma^{p,n}_{T,L}$ separately.
At the same time, the $\pi^0$-meson production amplitude on the deuteron is determined by the sum
of two amplitudes $M_{\pi^0 p}$ and $M_{\pi^0 n}$:
\begin{equation}
M_{\pi^0 p}:~~~(\gamma^\ast+p)+n\to (\pi^0+p)+n,~~~M_{\pi^0 n}:~~~(\gamma^\ast+n)+p\to (\pi^0+n)+p.
\end{equation}
Both these amplitudes have the same intermediate state $p+n+\pi^0$, so, we should take into account
the interference term: $|M_{\pi^0 p}+M_{\pi^0 n}|^2=|M_{\pi^0 n}|^2+|M_{\pi^0 n}|^2+{\cal I}$, where
${\cal I}=M^\ast_{\pi^0 p}M_{\pi^0 n}+M^\ast_{\pi^0 n}M_{\pi^0 p}$.
Then for correct calculation of the cross section of virtual photoabsorption
on the deuteron it is required phase fine-tuning of these amplitudes, what apparently can be done
by the authors of MAID \cite{MAID}. For crude estimate of $\sigma^d_{tot}$ we use an approximation
$\sigma^d_{tot}=\sigma^p_{tot}+\sigma^n_{tot}$.

\begin{figure}[t!]
\centering
\includegraphics[width=7.5 cm]{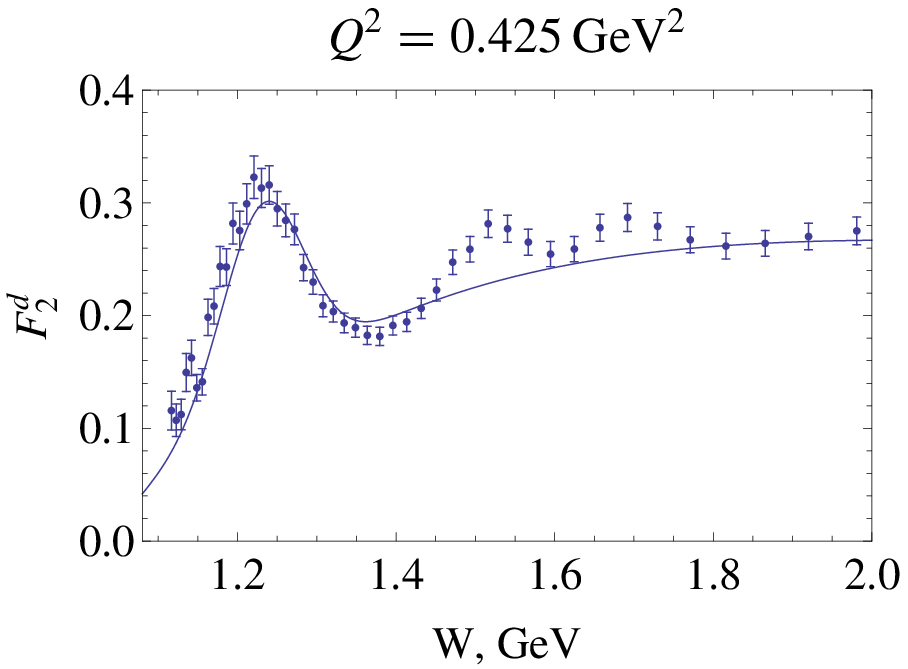}
\includegraphics[width=7.5 cm]{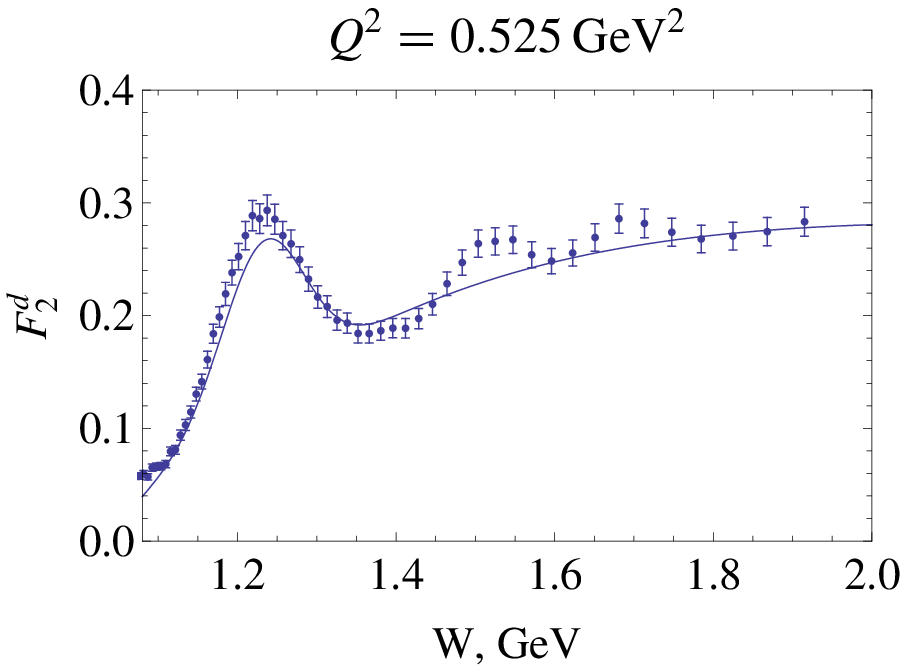}
\includegraphics[width=7.5 cm]{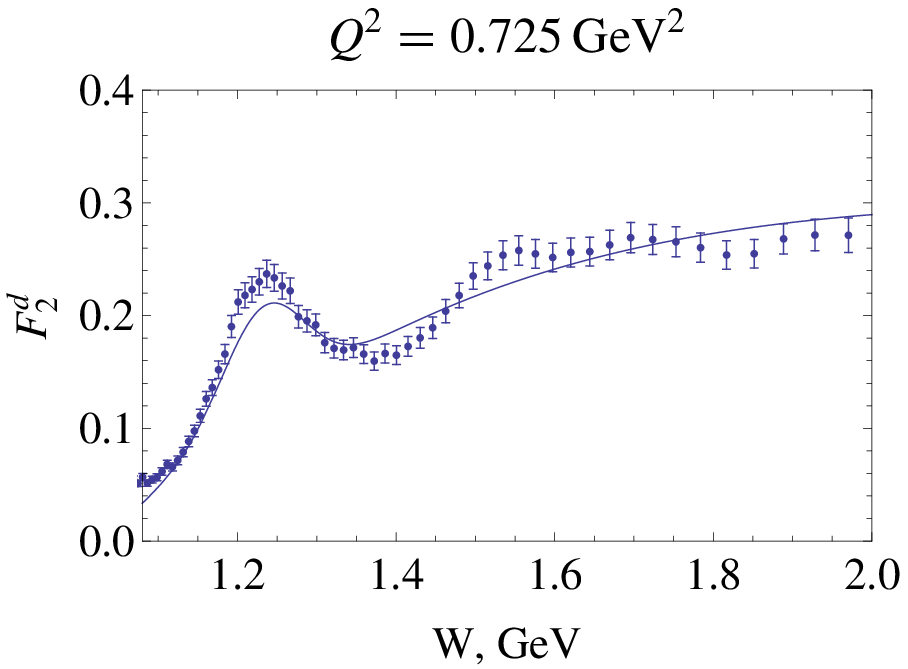}
\includegraphics[width=7.5 cm]{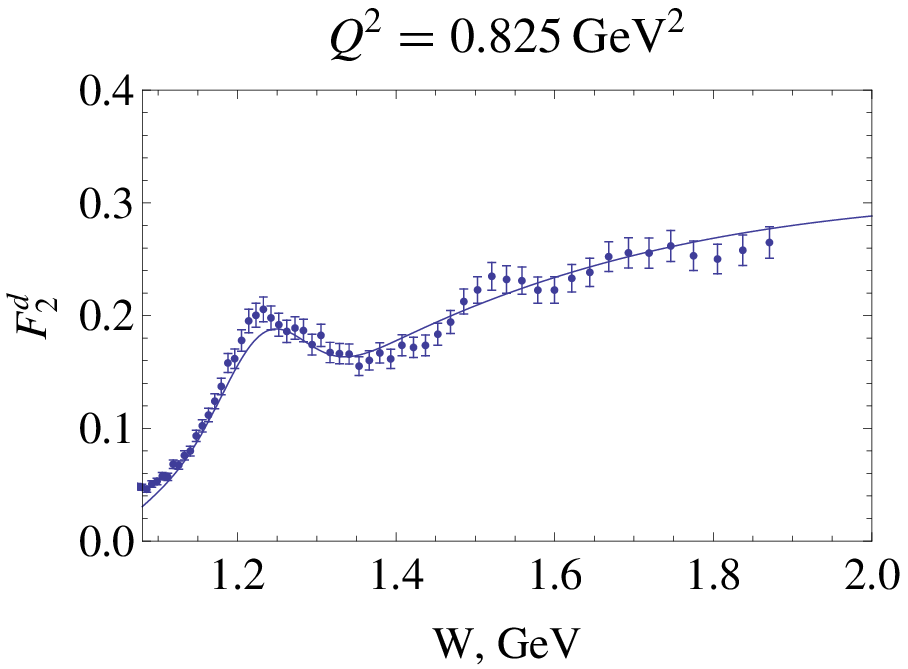}
\caption{Plots of deuteron structure function $F_2^d(W)$ per nucleon for $Q^2=0.425~GeV^2$, $Q^2=0.525~GeV^2$,
$Q^2=0.725~GeV^2$, $Q^2=0.825~GeV^2$ in the resonance region $W:~1.07\div 2.0$ GeV on the basis of \eqref{eq:ftot}.
Experimental points for $F_2^d$ are taken from \cite{osipenko}.}
\label{fig:sech}
\end{figure}

Another approach of the calculation in the resonance region which can be used for the improvement of the result obtained in MAID is based on the
experimental data on the deuteron structure function $F_2^d$ and their parameterization from \cite{amaudruz}:
\begin{equation}
F_2^d(W,Q^2)=[1-G^2(Q^2)][F^{dis}(W,Q^2)+F^{res}(W,Q^2)+F^{bg}(W,Q^2)],
\label{eq:ftot}
\end{equation}
where the contribution from a deep inelastic region was parameterize as
\begin{equation}
F^{dis}(W,Q^2)=\left[\frac{5}{18}\frac{3}{B(\eta_1,\eta_2+1)}x_w^{\eta_1}(1-x_w)^{\eta_2}+\frac{1}{3}\eta_3(1-x_w)^{\eta_4}\right]
S(W,Q^2),
\label{eq:fdis}
\end{equation}
\begin{displaymath}
\eta_i=\alpha_i+\beta_i \bar s,~~~\bar s=\ln\frac{\ln[(Q^2+m^2_a)/\Lambda^2]}
{\ln[(Q_0^2+m^2_a)/\Lambda^2]},
\end{displaymath}
\begin{displaymath}
x_w=\frac{Q^2+m_a^2}{2m_N\nu+m_b^2},~~~S(W,Q^2)=1-e^{-a(W-M_\Delta)^2/\Gamma^2},
\end{displaymath}
the contribution from the resonance region
\begin{equation}
F^{res}(W,Q^2)=\alpha_5^2G^{3/2}e^{-b(W-W_{thr})^2},
\label{eq:fres}
\end{equation}
and the background under the resonance region
\begin{equation}
F^{bg}(W,Q^2)=\alpha_6^2\xi G^{1/2}e^{-b(W-W_{thr})^2},
\label{eq:bg}
\end{equation}
The parameterization \eqref{eq:ftot} refers to the deuteron structure function $F_2^d$ normalized per nucleon.
Therefore, to obtain the full deuteron structure function $F_2^d$ it is necessary to multiply it by the number of nucleons.
The values of parameters in \eqref{eq:fdis}-\eqref{eq:bg} were taken from \cite{amaudruz}.
The parameterization \eqref{eq:ftot} which has a rather simple form for the work, was used by us earlier
in \cite{fm1} for the calculation of hydrogen-deuterium isotope shift. There are more complicated parameterizations
for $F_2^d$, such as in \cite{deut3}. As we consider the results of \cite{deut3} agree generally with \eqref{eq:ftot}.
The measurement of the deuteron structure function $F_2^d$ was carried out for decades by different experimental groups.
In the resonance region data on the structure function of the deuteron $F_2^d$ are presented in \cite{osipenko}.
Experimental results cover a broad kinematical range $0.375~GeV^2<Q^2<6~GeV^2$ and $0.1<x<1$ and have total
systematical errors typically near 5~$\%$. Despite the impressive achievements there is still region of kinematics
at small $Q^2$ where our knowledge of structure function $F_2^d$ remains poor.
In Fig.~\ref{fig:sech} we show a comparison of the parameterization \eqref{eq:ftot} of the deuteron structure function $F_2^d(W,Q^2)$
in the resonance region with existing experimental data from \cite{osipenko}.
Experimental data show that there is the "dip" region between the quasi-elastic peak and the $\Delta (1232)$ resonance.
We note that experimental points do not cover all relevant area $Q^2$, $W$ for estimate of deuteron
polarizability contribution with high accuracy. Despite the simplicity of the parameterization \eqref{eq:ftot} for $F_2^d$
existing experimental data agree well with it.
Numerical result for $\Delta E_{pol}^{LS}$ obtained using \eqref{eq:ftot} is presented in Table~\ref{tb1}.

In the nonresonance region there exists a parameterization for the function $F_2^d(Q^2,W)$ \cite{deut1} obtained on the basis of
experimental data on deep inelastic lepton-nucleon and lepton-deuteron scattering.
In this 23-parameter Regge-motivated model the structure function $F_2^d(x,Q^2)$ was expressed as a sum of the Pomeron $F_2^{\cal P}$ and the
Reggeon $F_2^{\cal R}$ term contributions for $W^2>4~GeV^2$, i.e., above the resonance region, and any $Q^2$ including the real photon point
($Q^2=0$):
\begin{equation}
\label{eq:f2d2}
F_2^d(x,Q^2)=\frac{Q^2}{Q^2+m_0^2}\left[F_2^{\cal R}(x,Q^2)+F_2^{\cal P}(x,Q^2)
\right],
\end{equation}
\begin{equation}
F_2^{\cal R}(x,Q^2)=C_{\cal R}(t)x_{\cal R}^{a_{\cal R}(t)}(1-x)^{b_{\cal R}
(t)},~~~
F_2^{\cal P}(x,Q^2)=C_{\cal P}(t)x_{\cal P}^{a_{\cal P}(t)}(1-x)^{b_{\cal P}
(t)},
\end{equation}
where $x$ is the Bjorken variable,
\begin{equation}
\frac{1}{x_{\cal R}}=1+\frac{W^2-m_N^2}{Q^2+m_{\cal R}^2},~~~
\frac{1}{x_{\cal P}}=1+\frac{W^2-m_N^2}{Q^2+m_{\cal P}^2},
\end{equation}
In the parameterization \eqref{eq:f2d2} the deuteron structure function is normalized per nucleon
as well as in \eqref{eq:ftot}.
Numerical values of the model parameters are presented in \cite{deut1}.

An accurate high statistics measurement of the ratio of the structure functions of the deuteron
and the proton, $F_2^d/F_2^p$, and the difference $R^d-R^p$ (R is the ratio of longitudinally to
transversely polarized virtual photon absorption cross sections), was obtained in deep inelastic
muon scattering. The values of $\Delta R=R^d-R^p$ are small. This is most significant at small $Q^2$.
For the structure function $R^p(Q^2,W)$ there was obtained the parameterization
in nonresonance region on the basis of experimental data \cite{Abe}. In the resonance region there are no experimental data
for the quantity $R^d(Q^2,W)$. In the most important part of the resonance region $\sigma_L\ll \sigma_T$. So,
to perform the numerical calculation of the correction $\Delta E^{LS}_{pol}$ we suppose that $R^d(Q^2,W)\approx 0$.

\begin{table}[!ht]
\caption{Hadronic deuteron polarizability correction to the Lamb shift $(2P-2S)$ of muonic deuterium.}
\label{tb1}
\bigskip
\begin{tabular}{|c|c|c|}     \hline
Method of the calculation &  MAID program &  Parametrization~\eqref{eq:ftot} of the  \\
$\Delta E_{pol}^{LS}, ~\mu eV$                     &   \cite{MAID}     &  deuteron function $F_2^d$ \cite{amaudruz}  \\   \hline
Contribution of $N\pi$ states                              &  30.88    &           \\   \cline{1-2}
Contribution of $N\eta$ states                              &  0.14    &       27.01         \\  \cline{1-2}
Contribution of K mesons                                &  0.16    &         \\    \cline{1-2}
Contribution of $N\pi\pi$ states                       &   3.00  &    \\   \hline
Nonresonance contribution                              &  1.93    &   1.93 \\    \hline
Contribution of the subtraction term                  &   -8.78     &  -8.78  \\   \hline
Total contribution                                  &  27.3    &  20.2   \\    \hline
\end{tabular}
\end{table}

There exists a number of theoretical uncertainties connected with quantities entering in the correction \eqref{eq:7}.
In the improved isobar model \cite{UIM1,UIM2} containing 15 resonances, we can omit theoretical error which arises due to the
insertion of other high-lying nucleon resonances. On our sight the main theoretical error is closely related with the calculation of the helicity
amplitudes $A_{1/2}(Q^2)$, $A_{3/2}(Q^2)$, $S_{1/2}(Q^2)$ in the quark model based on the oscillator potential \cite{CL}.
Only systematical experimental data for the helicity amplitudes of the photoproduction on the nucleons $A_{1/2}(0)$, $A_{3/2}(0)$ are known
with sufficiently high accuracy to the present \cite{PDG}. In the case of amplitudes for the electroproduction of the nucleon resonances
experimental data contain only their values at several points $Q^2$. So, we have no consistent check for the predictions of the oscillator model.
Possible theoretical uncertainty connected with the calculation of amplitudes $A_{1/2}(Q^2)$, $A_{3/2}(Q^2)$, $S_{1/2}(Q^2)$ with the account
of relativistic corrections may be at least 10~$\%$. Then the theoretical error
for the correction \eqref{eq:7} in the resonance region up to 20~$\%$.
There is theoretical uncertainty in the contribution due to two-pion nonresonance processes which are presented above.
The error in this case may not be less than 30 $\%$.
The essential part of theoretical error is connected with the subtraction term because deuteron magnetic
polarizability is known with a precision 30~$\%$. Moreover, since there are no experimental data about $k$-dependence
of $C_1(0,k^2)$ in \eqref{eq:2} we assume, following to \cite{kp1999} that it is described by the function~\eqref{eq:5}.
The form of function $\beta_M(k^2)$
in~\eqref{eq:5} also provides additional theoretical error. But since it can not be reasonably estimated, we estimate the
total theoretical uncertainty of the subtraction term in~\eqref{eq:7} of the parameter $\beta^d_M(0)$ in $\pm 2.6~\mu eV$
for 2S-state or near 10~$\%$ of total result.
As we discuss above the most significant part of the error in the calculation on the basis of the MAID
is related with the interference of pion production amplitudes. Numerical estimate of the maximum and minimum values
of production cross sections shows that theoretical error is equal approximately to $\pm 8~\mu eV$ or 25~$\%$ for 2S-state.
Total theoretical error in the MAID calculation amounts to 45~$\%$. An estimate of theoretical error at the calculation
on the basis of experimental data can de derived from systematical and statistical errors of the deuteron function $F_2^d$
measurements.
It does not exceed in this case 10~$\%$. But the parameterization \eqref{eq:ftot}, which comprises essentially only the
contribution of $\Delta$-isobar is not always a good description of other resonances (see Fig.~\ref{fig:sech}).
Therefore, a reasonable estimate of the maximum error increases to 20 $\%$ when using \eqref{eq:ftot}.
It is necessary to mention that there are no experimental data in the region
of small values of $Q^2$, which may eventually lead to much greater error.
The obtained central value of hadronic deuteron polarizability contribution to the Lamb shift
$(2P-2S)$ in muonic deuterium presented in Table~\ref{tb1} is in agreement with previous calculations in \cite{carlson,kp2015,bacca}.
In a discussion of hadronic contribution to the deuteron polarizability in \cite{CREMA2015,carlson,kp2015,bacca} it was
suggested that this contribution is approximately equal to the sum of contributions of the proton and neutron forming a deuteron:
\begin{equation}
\label{eq:bacca}
\delta_{pol}^{hadr}(\mu X)=Z^3(Z+N)[m_r(\mu X)/m_r(\mu H)]^3\delta_{pol}^{hadr}(\mu H).
\end{equation}
Numerical value of hadronic contribution to the Lamb shift in muonic deuterium
0.028 meV was presented in Table~IV of recent summary paper \cite{CREMA2015}.
In our calculation of this contribution by means of MAID we consider that the deuteron structure function $F_2^d$
is determined by \eqref{eq:f2pf2n}, what naturally leads to the value of the contribution close to the result \eqref{eq:bacca}.
Numerous experimental data, including the above-mentioned \cite{deut1,deut2,deut3} show that the approximate equation $F_2^d\approx F_2^p$,
where $F_2^d$ is determined as the deuteron structure function per nucleon,
sufficiently accurately holds in different kinematic regions.
Our other calculation of hadronic polarizability contribution on the basis of experimental data for $F_2^d$ \eqref{eq:ftot} 
gives smaller result 20.2 $\mu eV$.
The difference of this result from the value of 0.028 meV \cite{CREMA2015,carlson} may be due to the subtraction term.
The subtraction term which was presented in Table~I of \cite{carlson} as a separate line, has essentially different value
because it is determined by total deuteron magnetic polarizability value $\beta_M^d=0.072(5)~fm^3$ (compare with our
value in \eqref{eq:4}).
It should be noted that there is a difference between the deuteron structure function $F_2^d$, built within MAID
and experimental data from \cite{amaudruz,osipenko} what is particularly evident in the area of the $\Delta$-isobar.
It can be explained by the influence of the interference terms in the amplitude of the Compton scattering on the deuteron,
which can lead to a decrease in the value of the function $F_2^d$ in the area of the $\Delta$-isobar and its increase in the
region of higher-lying resonances. The difference in the used deuteron structure functions MAID and \eqref{eq:ftot} 
affects the final results, which are presented in Table~\ref{tb1}.
The simplest way to understand qualitatively our result is the following. The proton
and neutron in the deuteron form a loosely bound system, so, we can consider the virtual Compton scattering separately on
the proton and neutron neglecting nuclear effects. The cross sections of virtual photo-absorption on the proton and
neutron approximately coincide. The calculation of proton polarizability contribution was carried out in \cite{m1}. Thus,
neglecting interference terms in the production cross sections we have crude estimate of the hadronic deuteron polarizability
contribution $\Delta E_{pol}^{LS}(deuteron)\approx 2\Delta E_{pol}^{LS}(proton)$. An account of reduced mass dependence
slightly changes this result. Nuclear shadowing will change the approximation $F_2^d\approx F_2^p$ (the neutron structure
function $F_2^n\approx F_2^p$), but the ratio $F_2^d/F_2^p$
can be slightly less than or greater than~1 for different kinematical regions. In this sense shadowing can be regarded as a part of
known EMC effect \cite{emc}. New more precise experimental data on the deuteron structure function $F_2^d$ especially in the resonance
region at small values of $Q^2$ could improve the calculation of hadronic deuteron polarizability contribution to the Lamb shift
in muonic deuterium.

\begin{acknowledgments}
We are grateful to M.~Gorshtein, M.~Osipenko and R.~Pohl for useful communications, critical remarks and discussion
the problem of this paper.
The work is supported by the Russian Foundation for Basic Research (grant 16-02-00554),
the Ministry of Education and Science of Russia under Competitiveness Enhancement Program 2013-2020
and grant No.~1394.
\end{acknowledgments}

\end{document}